\RequirePackage{fixltx2e}
\documentclass[aip,jcp,reprint,showpacs,singlecolumn]{revtex4-1}

\usepackage{dcolumn}
\usepackage{placeins}
\usepackage{epsf}
\usepackage{subfigure}
\usepackage{epsfig,amsopn}
\usepackage{amssymb,graphicx}
\usepackage{amsmath,amsfonts}
\usepackage{graphicx}
\usepackage{dcolumn}
\usepackage{bm}
\usepackage{ifpdf}
\usepackage{natbib}
\usepackage{wasysym}
\usepackage[byname]{smartref}
\usepackage{color}
\usepackage{lipsum}
\usepackage{nicefrac}
\usepackage[breaklinks, colorlinks, linkcolor=blue, citecolor=blue, urlcolor=blue]{hyperref}
\usepackage[table]{xcolor}
\usepackage{multirow}
\usepackage{braket}
\usepackage{float}
\usepackage{comment}
\usepackage{enumerate}
\usepackage{enumerate}
\usepackage{tabularx}

\newcommand{\beq}{\begin{equation}}
\newcommand{\eeq}{\end{equation}}

\newcommand{\bea}{\begin{eqnarray}}
\newcommand{\eea}{\end{eqnarray}}

\begin{document}

\title{
Controlling charge transport mechanisms in molecular junctions: Distilling thermally-induced hopping from coherent-resonant conduction}

\author{Hyehwang Kim}
\affiliation{Chemical Physics Theory Group, Department of Chemistry, University of Toronto,
80 St. George Street Toronto, Ontario, Canada M5S 3H6}

\author{Dvira Segal}
\email{dsegal@chem.utoronto.ca}
\affiliation{Chemical Physics Theory Group, Department of Chemistry, University of Toronto,
80 St. George Street Toronto, Ontario, Canada M5S 3H6}

\date{\today}
\begin{abstract}
The electrical conductance of molecular junctions may
strongly depend on the temperature, and  weakly on molecular length, under two distinct mechanisms:
phase-coherent resonant conduction, with charges proceeding via delocalized molecular orbitals, 
and incoherent thermally-assisted multi-step hopping.
While in the case of coherent conduction the temperature dependence arises from the broadening of the Fermi distribution 
in the metal electrodes, in the latter case it corresponds to electron-vibration interaction effects on the junction.
With the objective to distill the thermally-activated hopping component,
thus expose intrinsic electron-vibration interaction phenomena on the junction,
we suggest the design of molecular junctions with ``spacers", extended anchoring groups that act to filter out
phase-coherent resonant electrons.
Specifically, we study the electrical conductance of 
fixed-gap and variable-gap junctions that include a tunneling block, with 
spacers at the boundaries.
Using numerical simulations and analytical considerations, we demonstrate that in our design, resonant conduction
is suppressed. 
As a result, the electrical conductance is dominated by two (rather than three) mechanisms: superexchange (deep tunneling), and multi-step thermally-induced hopping.
We further exemplify our analysis on DNA junctions with an A:T block serving as a tunneling barrier. Here,
we show that the  electrical conductance is insensitive to the number of G:C base-pairs at the boundaries.
This indicates that the tunneling-to-hopping crossover revealed in such sequences 
truly corresponds to the properties of the A:T barrier.
\end{abstract}

\maketitle


\section{Introduction}
\label{Sintro}

Electrical conductance measurements in single molecules and self-assembled 
monolayers often reveal {\it three} competing transport mechanisms \cite{nitzan,scheer}:
(i) phase-coherent off-resonance tunneling, also referred to as ``superexchange" or deep tunneling, (ii) 
phase-coherent on-resonance tunneling, also referred to as ballistic motion, 
and (iii) incoherent multi-step hopping, which may be thermally activated. 

These three mechanisms are traditionally identified through the examination of the 
length and temperature dependence of the electrical conductance $G$.
Considering a molecular bridge setup, with the molecular orbitals situated away (beyond thermal energy) from the Fermi energy of the leads, the following characteristics are revealed:
In the deep tunneling mechanism, 
$G$ does not depend on temperature as it is proportional to the transmission probability of electrons at the Fermi energy. 
It decreases exponentially with molecular length $N$, $G\propto \exp(-\alpha N)$, therefore being ineffective in long molecules.  
Here $\alpha$ is the decay constant, depending on the bridge energetics \cite{nitzan}.
In contrast, phase-coherent {\it resonant} conduction is largely insensitive to the molecular length, $G\sim  N^0$, 
as electrons proceed through delocalized molecular orbitals.
Furthermore, in this limit $G$ displays a strong dependence on temperature, essentially, a thermally activated behavior 
following the Fermi distribution of the electrodes.
The third mechanism, incoherent multi-step hopping, is facilitated in 
long molecules---once the transit time of electrons through the molecular bridge is long enough---allowing them to 
interact with ``environmental"  (internal or external) degrees of freedom, e.g., phonons. 
This process is thermally induced, and it is typically
characterized by a thermal activation-Arrhenius factor. Since the interaction of electrons with the nuclei
brings about electronic energy dissipation,
in this limit the resistance typically grows linearly
with molecular size as $G\propto (aN+b)^{-1}$
($a$ and $b$ depend on the bridge energetics, temperature, thermal effects) \cite{Segal00},
eventually approaching an ohmic behavior.

The competition between superexchange tunneling and thermally-induced hopping transport has been 
the subject of many investigations, examined in 
 different types of molecules, such as conjugated organic molecules
 \cite{selzer1,selzer2,selzer3,Tao10,Frisbie1,Frisbie2, Frisbie3,Tada} and biomolecules  
\cite{Bixon,Barton10,GieseE,GieseR,Tao04,Tao16,Cahen14,Gray,NijhuisF}. 
Observations of nearly length-independent conduction \cite{Nichols,Lin,Sharif}, or temperature dependent conduction
 which can be associated to the thermal broadening of the Fermi functions
\cite{Zant, Nichols,Nijhuis16,NijhuisDal}, point to phase-coherent resonant 
transmission as the dominant transport mechanism.

Certainly, many experiments do not fit into this simple classification,
exposing  distinct transport behavior.
Such ``unconventional"  results are argued
to be linked to e.g., conformation changes  \cite{latha14},
transitions in binding geometry as the molecular length increases \cite{TaoI}, as well as
structural changes in the electrode---induced by the temperature \cite{lathaT}.
Such effects are beyond the scope of our work.

From the technological point of view,
designing molecular wires that support length-independent resonant transmission is beneficial 
for some applications. In such cases,
the conductance essentially evinces on (i) 
the quality of the molecule-metal contact, and (ii) the energy of the relevant delocalized molecular orbitals 
relative to the chemical potential of the leads.
However, if one's objective is to learn about electron-vibration interaction effects (inelastic scattering,
heat generation, phonon transport, phonon damping, phonon cooling), that are  
intrinsic to the molecular entity, one should aim to reduce the ``irrelevant" ballistic contribution, 
so as to observe signatures of electron-nuclei interaction effects, in particular,
the tunneling-to-hopping crossover.
This challenge is especially relevant to relatively short molecular junctions 
in which the three mechanisms discussed above can show up simultaneously, 
to confound the identification of the dominant transport mechanism
\cite{selzer1,selzer2,selzer3,Tao10,Frisbie1, Frisbie2, Tada,Tao04, Tao16,Zant,Nichols, Lin,Nijhuis16}.

\begin{figure*}[ht]
\vspace{0mm} \hspace{3mm}
{\hbox{\epsfxsize=180mm \epsffile{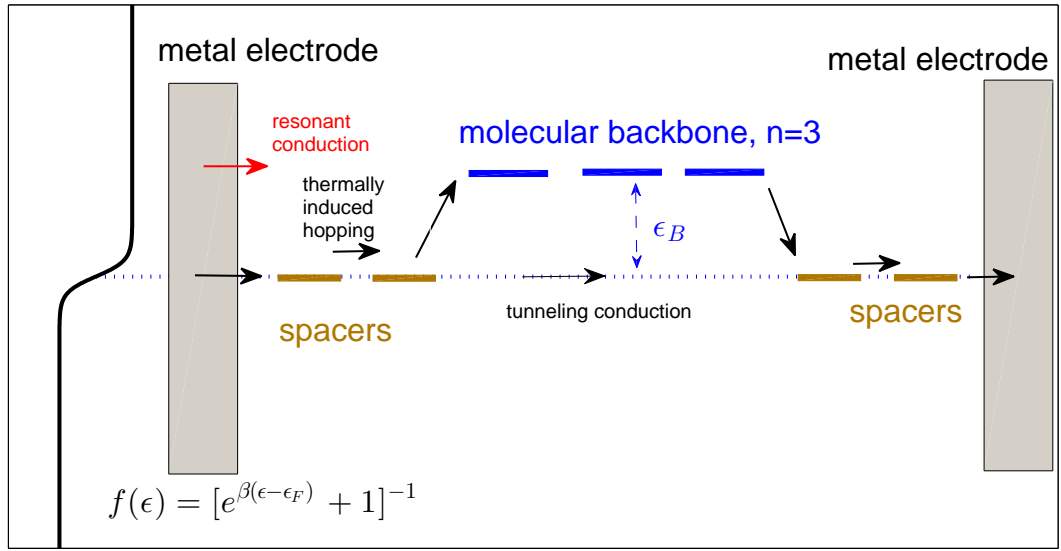}}}
\vspace{-15mm}
\caption{Design principle of molecular junctions with suppressed coherent-resonant conduction. 
The structure includes  repeating molecular units, with spacers (possibly made of repeating units) at the boundaries.
The spacers are chosen so as their energies are set close to the Fermi energy (dotted line).
The Fermi function at the electrodes is depicted at the left end. In the absence of spacers, electrons arriving at the tail of the Fermi function
 (red arrow) cross the junction through delocalized orbitals. However, spacers act as a tunneling barrier for these 
electrons, suppressing this contribution.
 In contrast, electrons incoming at the Fermi energy (black arrow)
 resonantly transmit through the spacers,  and cross the molecular barrier via a deep-tunneling
mechanism, or a hopping mechanism---after being thermally-activated.
}
\label{diag1D}
\end{figure*}

In this paper, we propose  
a simple design for molecular junctions, with the objective to suppress
phase-coherent resonant conduction.
In our setup, electrons reach the molecular entity `concentrated' at the Fermi energy---with a minimal thermal tail.
This is achieved by connecting molecules of interest to the metal leads via extended anchoring-linker groups 
(below referred to as `spacers'), making it difficult for electrons arriving off the Fermi energy to cross the junction,
see Fig. \ref{diag1D}.
As a result, the electrical conductance in our setup shows a pure
tunneling-to-hopping crossover as a function of both length and temperature, 
free from the (confusing) coherent-resonant contribution. 

To model the interaction of conducting charges with nuclear degrees of freedom (and other scattering events)
we use  the Landauer-B\"uttiker probe (LBP) technique \cite{Buttiker1,Buttiker2}.
In this approach, incoherent elastic and inelastic scattering effects are introduced 
by augmenting the non-interacting electronic Hamiltonian with probe terminals---through which electrons 
lose their phase memory and exchange energy. 
The probes mimic the action of
an ``environment", a collection of intra-molecular and inter-molecular degrees of freedom 
(vibrational, electronic, impurities). Note that the source-drain
metal electrodes are not considered here as part of the ``thermal environment".

The LBP technique was originally introduced to model decoherence effects in mesoscopic systems \cite{Buttiker1,Buttiker2}.
More recently, it was applied to investigate e.g. electronic
conduction in organic and biological molecular junctions \cite{Nozaki1,Nozaki2,Chen-Ratner,WaldeckF,spinPRL,Qi}.
Particularly, it was recently demonstrated in Refs. \cite{Kilgour1, Roman} that the LBP
method can capture different transport regimes in molecular junctions: deep tunneling conduction,
ballistic motion, incoherent hopping, 
as well as an intermediate quantum coherent-incoherent regime \cite{William}. 
In Refs. \cite{Kilgour2,Kilgour3}, we further used the LBP technique
to simulate high-bias voltage effects, specifically, the role of environmental interactions
on the operation of a molecular junction as a diode.

Making use of the LBP technique, we solve a simple analytical model and perform simulations, 
to demonstrate that in our design (a molecule with spacers at the boundaries)
we filter out resonant electrons. 
After exemplifying this principle on a 1-dimensional (1D) chain, we simulate the electrical conductance 
of DNA molecules with an A:T segment---acting as a tunneling barrier---using G:C bases as spacers. 
Here, A, G, C and T are the adenine, guanine, cytosine and thymine bases, respectively.
We demonstrate that the spacers do not affect the tunneling-to-hopping crossover, 
which thus reflects intrinsic properties of the (A:T)$_n$ barrier.

The paper is organized as follows. In Sec. \ref{design}, we describe our setup.
We outline how we incorporate environmental interactions via the probe method in Sec. \ref{Hmethod}.
Numerical and analytical results from a 1D model are included in Sec. \ref{Result1D}.
In Sec. \ref{ResultDNA}, we exemplify our design on DNA sequences with an (A:T)$_n$ barrier.
We summarize our work in Sec. \ref{Summ}.


\section{Design Principle}
\label{design}

Structures of interest include repeating units, (m)$_n$.
The entity `m' could be a conjugated monomer \cite{Frisbie1,Frisbie2}, or a nucleotide base \cite{Tao16}. 
Our objective is to reveal charge transport mechanisms through this molecule, specifically, 
signatures of electron-vibration interaction effects.

In a typical experiment (Junction I),
a family of molecules is synthesized, by increasing the number of repeating molecular units, 
L--(m)$_n$--R with $n=1,2,3,...$.
The molecule is coupled to the L and R electrodes at the edges, see Table I. 
The linearity of such molecules, combined with the assumption of weak structural and dynamical disorder, implies that
resonant conduction, rather than multi-step hopping, may dominate its conduction (even in long molecules).
As mentioned in the Introduction, we view this as an obstacle: The contribution of resonant electrons 
outweighs inelastic multi-step hopping processes, the result
of rich electron-phonon interaction effects.

To alleviate this problem, we attach spacers to the left and right of the
molecular backbone. Specifically, we consider here spacers made of repeating units, (s)$_k$.
The overall molecular junction now takes the form
 L--(s)$_k$(m)$_n$(s)$_k$--R.  
The two components, `m' and `s', have different relevant energies:
The HOMO/LUMO states of `m' are placed away from the Fermi energy, while the `s' entity is chosen such that its relevant 
levels are aligned close to the Fermi level.

Fig. \ref{diag1D} represents our design.
Why does this structure lead to the suppression of resonant conduction? 
Incoming electrons around the Fermi energy (black arrow) can readily cross the spacers, to reach the molecular barrier, 
then tunnel through it, or hop through assisted by the nuclei motion.
In contrast, electrons arriving at the tail of the Fermi function, specifically, in resonance with
molecular states at $\epsilon_B$ (red arrow),
poorly cross the spacers, which collectively act as a tunneling barrier for high-energy carriers.
A different way to explain this design is to note that spacers modify the density of states (DOS) of the metal electrodes, 
to enhance the DOS around the Fermi energy, thus promote conduction from this state.

Optimally, spacers should not impact the intrinsic properties of the 
molecular core, (m)$_n$. Thus, one can consider different types of junctions with spacers: 
molecules  with a  fixed number of spacers at the boundaries (Junction II), or molecules
with a fixed total number of units (Junction III), while replacing the center of the molecule by the `m' monomers.

In Table I we list three families of molecules examined in this work.
Junction I, without spacers, and Junctions II and III of fixed and variable length, which differ in the number of spacers included.
Below, we compare the properties of Junction II and Junction III, and confirm that they (almost) 
identically operate, supporting our proposition that spacers act to suppress the resonant current, while minimally affecting
the transport properties of the original molecule (m)$_n$.

The fact that spacers (anchors or linkers) can modify the transport behavior of molecular junctions is certainly well known, 
see for example Refs. \cite{Cahen14,Hihath14,lathaRev}.
Particularly interesting is Ref. \cite{switch}, where the incorporation of anchor (methylene) groups
lead to the reduction of the electrode-molecule coupling, a key factor in the realization of
stable and reversible photoswitches.
These studies and others had emphasized the role of anchors on the contact resistance and energy level alignment.
Here, our focus is on the role of spacers in suppressing an undesirable transport mechanism.
In particular, our contribution here is in demonstrating that, in an optimal design, spacers very effectively 
filter out resonant conduction, to refine the contribution of nuclei-assisted charge conduction.
In this respect, we show that our construction is robust, with Junction II and Junction III supporting a similar behavior.

\vspace{5mm}
\begin{table*}[htbp]
\hspace{10.4mm } Table I:   {\bf  Molecular junctions examined in Sec. \ref{Result1D}.\\
L and R represent the electrodes, with spacer (s) and molecular (m) units.  }\\
\begin{tabular}{c c c c}
\hline
molecular length  \hspace{1mm}   \vline &   \hspace{5mm} Junction I  \hspace{16mm} \vline                  &         \hspace{5mm} Junction II  \hspace{16mm} \vline                  &  Junction III  \\ [0.5ex]
  (\# of m units)   \hspace{4mm} \vline &    (total length $N$=1 to 11)    \hspace{1mm} \vline &                                             (total length $N$=5 to 15)    \hspace{2.3mm} \vline  &       (total length $N=13$)    \\ [0.5ex]
\hline
\hline
\hspace{12mm }1 \hspace{11mm} \vline &  L--(m)$_1$--R    &   L--(s)$_2$ (m)$_1$(s)$_2$--R  &    L--(s)$_6$(m)$_1$(s)$_6$--R  \\
\hline 
\hspace{12mm }3 \hspace{11mm} \vline &  L--(m)$_3$--R &    L--(s)$_2$(m)$_3$(s)$_2$--R  & L--(s)$_5$(m)$_3$(s)$_5$--R  \\
\hline
\hspace{12mm }5 \hspace{11mm} \vline &  L--(m)$_5$--R  &     L--(s)$_2$(m)$_5$(s)$_2$--R   &  L--(s)$_4$(m)$_5$(s)$_4$--R \\
\hline
\hspace{12mm }7 \hspace{11mm} \vline &  L--(m)$_7$--R   &   L--(s)$_2$(m)$_7$(s)$_2$--R  & L--(s)$_3$(m)$_7$(s)$_3$--R \\
\hline
\hspace{12mm }9 \hspace{11mm} \vline &  L--(m)$_9$--R    &   L--(s)$_2$(m)$_9$(s)$_2$--R  & L--(s)$_2$(m)$_9$(s)$_2$--R \\
\hline
\hspace{11mm }11 \hspace{10.5mm} \vline &  L--(m)$_{11}$--R  &  L--(s)$_2$(m)$_{11}$(s)$_2$--R  &  L--(s)$_1$(m)$_{11}$(s)$_1$--R \\
\hline
\end{tabular}
\label{table:seq}
\end{table*}


\section{Hamiltonian and Method}
\label{Hmethod}

We model the molecular junction by a tight-binding Hamiltonian. Each unit in the molecule is 
described by a single electronic site (level);
for an interesting discussion over the connection between this description, and an atomic picture, see Ref. \cite{Hsu}.
The total Hamiltonian reads
\bea
\hat H&=&\hat H_M+\hat H_{L,R} + \hat H_C,
\nonumber\\
\hat H_M &=& \sum_{n}\epsilon_n \hat c_n^{\dagger} \hat c_n 
+\sum_{n} \left(v_{n,n+1} \hat c_n^{\dagger}c_{n+1} + h.c.\right).
\nonumber\\
\hat H_{L,R}&=& \sum_{l}\epsilon_l \hat c_l^{\dagger}c_l  + \sum_{r}\epsilon_r \hat c_r^{\dagger}c_r
\nonumber\\
\hat H_C&=& \sum_{l}g_{L,l} \hat c_{1}^{\dagger}c_l  + \sum_{r}g_{R,r} \hat c_N^{\dagger}c_r.
\label{eq:H}
\eea
Here, $\hat c_{n,l,r}^{\dagger}$ ($\hat c_{n,l,r}$) are fermionic creation (annihilation) operators
of electrons on each site $n$, and in the metallic states.

In Junction I, L--(m)$_{n}$--R,  the
repeating units correspond to the molecular backbone. We set
$\epsilon_B=\epsilon_n$ and $v=v_{n,n+1}$.
To study the tunneling-to-hopping crossover, 
we increase the molecular length by adding units to the chain, see e.g. 
\cite{Frisbie1,Frisbie2,Tao16}.
In Junction II and III,  L--(s)$_{k}$(m)$_{n}$(s)$_k$--R, 
the energy of a unit spacer `s' is
$\epsilon_s$, and it is set at the Fermi energy. The molecular unit `m' takes the energy 
$\epsilon_B$ (as in Junction I). For simplicity, the tunneling elements are made uniform throughout, $v=v_{n,n+1}$. 

Using the LBP method, we include incoherent effects on the junction by attaching B\"uttiker probes
to each `m' and `s' site.
We outline next the LBP approach. For details, 
see e.g. Refs. \cite{Kilgour1,Kilgour2,Kilgour3,Roman,William}.

The total-net charge current, leaving the $\alpha$ (L,R, probes) contact, is written as (per spin)
\bea
I_{\alpha}=\frac{e}{h}\sum_{\alpha'} \int_{-\infty}^{\infty} \mathcal T_{\alpha,\alpha'}(\epsilon) \left[ f_{\alpha}(\epsilon)-f_{\alpha'}(\epsilon)\right] d\epsilon.
\label{eq:curr}
\eea
Here, $f_{\alpha}(\epsilon)=\left[ e^{\beta (\epsilon-\mu_{\alpha})}+1\right]^{-1}$ is the Fermi distribution function.
It is defined in terms of the temperature $k_BT=\beta^{-1}$, which is set uniform across 
the junction, and the chemical potential $\mu_{\alpha}$. The chemical potentials of the $L$ and $R$ electrodes are given;
the parameters of the probes are determined from the probe condition, described below.

The transmission function is obtained from the retarded and advanced 
Green's function and the metal-molecule hybridization matrices, as
\bea
\mathcal T_{\alpha,\alpha'}(\epsilon)= {\rm Tr}\left[ \hat \Gamma_{\alpha}(\epsilon) \hat G^{r}(\epsilon)
\hat \Gamma_{\alpha'}(\epsilon) \hat G^{a}(\epsilon)\right].
\eea
The retarded Green's function is given by
\bea
\hat G^r(\epsilon)=\left[ \epsilon- \hat H_M + i \hat \Gamma(\epsilon)/2\right]^{-1},
\eea
$\hat \Gamma(\epsilon)=\hat \Gamma_L(\epsilon) + \hat \Gamma_R(\epsilon) +\sum_p\hat \Gamma_p(\epsilon)$, including
the hybridization matrices to the left, right, and probe terminals.
In our geometry, these matrices comprise a single nonzero value,
\bea
[\Gamma_{L}]_{1,1}= \gamma_L, \,\,\, 
[\Gamma_{R}]_{N,N}= \gamma_R,\,\,\,
[\Gamma_{p}]_{p,p}= \gamma_p.
\eea
Here, $N$ corresponds to the last site of the chain, coupled to the $R$ terminal.
$\gamma_{\nu=L,R}(\epsilon)=\sum_{j \in \nu} |g_{\nu,j}|^2\delta(\epsilon-\epsilon_j)$ 
is the metal-molecule coupling (hybridization) energy. $\hbar/\gamma_p$ is the time scale for all incoherent effect
including phase loss, momentum and energy exchange. 
In our simulations below we take the parameters $\gamma_{L,R,p}$ to be constant, energy
independent.

Working in the linear response regime with $\Delta\mu\equiv (\mu_L-\mu_R)\ll k_BT,|\epsilon_B-\epsilon_F|,\gamma_{L,R}$,
the chemical potentials of the probes are determined from the so-called voltage probe condition: We enforce charge conservation 
between the source $L$ and drain $R$ by demanding that the charge current flowing {\it to each probe} is null. This set of 
conditions translate into linear equations for the chemical potentials of the probes $\mu_p$. The solution  is used in Eq. (\ref{eq:curr}),
to calculate the net source-drain current $I_L$. 

\begin{figure}[ht]
\vspace{0mm} \hspace{3mm}
{\hbox{\epsfxsize=85mm \epsffile{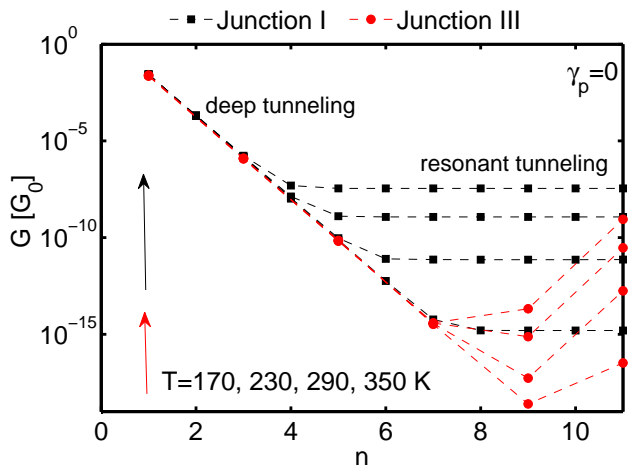}}}
\caption{Length dependent coherent conduction.
Junction I ($\square$), Junction III ($\circ$) 
with $\epsilon_B=0.6$ eV,  $\epsilon_s=0$, $v=0.05$ eV, $\gamma_{L,R}=0.1$ eV, $\gamma_p=0$.
}
\label{Fig1}
\end{figure}

\begin{figure}[ht]
\vspace{0mm} \hspace{3mm}
{\hbox{\epsfxsize=90mm \epsffile{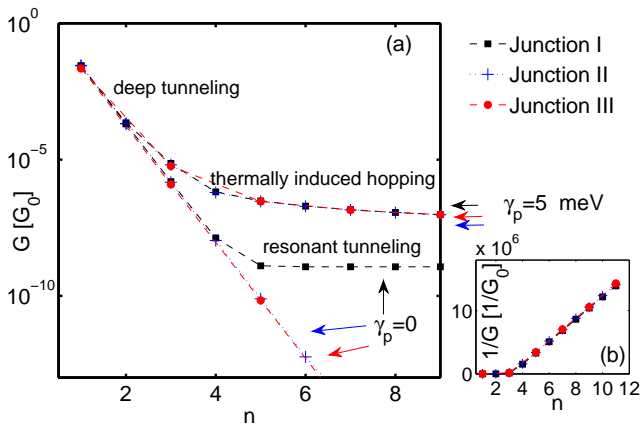}}}
\caption{
(a) Length dependent conductance at  $\gamma_p=0$ and $\gamma_p=5$ meV. 
Junction I ($\square$),  Junction II  ($+$), and Junction III ($\circ$).
We use the same parameters as in Fig. \ref{Fig1} with $T=290$ K. 
In panel (b) we confirm the onset of the hopping-ohmic  behavior in long junctions.
}
\label{Fig2}
\end{figure}

\section{Proof of principle: 1D junction}
\label{Result1D}

\subsection{Simulations}

We perform the LBP simulations by assuming wide, structure-less metallic bands.
We set the Fermi energy at $\epsilon_F=0$. 
Our simulations enforce the linear response condition (small voltage bias) through the linearized LBP equations \cite{Kilgour1}.
Below we plot the electrical conductance
$G\equiv I_L/V$, $\Delta \mu=eV$, in units of $G_0=e^2/h$. 

The length dependence of the electrical conductance in Junctions I-III is displayed in Figs. \ref{Fig1}-\ref{Fig2}.
In Figure \ref{Fig1}, we examine the coherent limit with $\gamma_p=0$. 
We demonstrate that in Junction I, the conductance displays a superexchange-to-resonant 
tunneling crossover at room temperature around $n=4-5$ molecular units. 
In contrast, in Junction III (and similarly, in Junction II--not shown)
we manage to significantly reduce the resonant contribution, 
with only off-resonant tunneling electrons contributing in molecules $n=1-7$ units long.
Beyond that, off-resonant tunneling is negligible, and residual, resonant conduction takes over 
at high enough temperatures.

We study the role of electron-vibration interaction effects, emulated here by the probes,
in Fig. \ref{Fig2}. We find 
that when $\gamma_p\neq0$, the conductance crosses over from a strong exponential decay to the slow function
 $G\propto (aN+b)^{-1}$, which characterizes the hopping regime \cite{Segal00}, see panel (b). Here, $a$ and $b$
are constants, expressed in terms of the junction's parameters \cite{Kilgour1}.
%
Fig. \ref{Fig2} clearly demonstrates the advantage of a setup with spacers, Junction II and III, over Junction I.
In the latter case, the magnitude and the behavior of the conductance is 
quite similar with and without environmental effects,
given the substantial contribution of resonant electrons.
This correspondence poses a significant challenge in experiments,
confounding the determination of transport mechanisms.
In contrast, in Junctions II and III, marked differences show up between the coherent and incoherent regimes:
In the coherent case ($\gamma_p=0$) the superexchange mechanism prevails, while when $\gamma_p\neq 0$,
hopping conduction dominates in long chains, approaching an ohmic resistance. 

It is also important to note that
the three setups, Junction I, II, and III,
perform identically in the hopping regime (overlapping lines in Fig. \ref{Fig2}).
This observation critically supports our assertion that, in an optimal design,  
spacers filter out resonant electrons, but leave intact the intrinsic electronic 
properties of the molecular block, (m)$_n$.

\begin{figure}[ht]
\vspace{0mm} \hspace{3mm}
{\hbox{\epsfxsize=85mm \epsffile{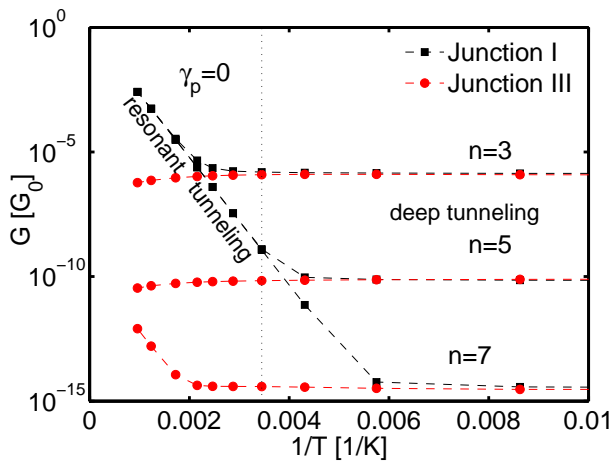}}}
\caption{Temperature dependence  of the conductance in the coherent limit $\gamma_p=0$.
Junction I ($\square$) and Junction III ($\circ$),
with $\epsilon_B=0.6$ eV, $\epsilon_s=0$, $v=0.05$ eV, $\gamma_{L,R}=0.1$ eV, $n=3,5,7$.
The dotted line marks room temperature.
}
\label{Fig3}
\end{figure}
\begin{figure*}[ht]
\vspace{0mm} \hspace{3mm}
{\hbox{\epsfxsize=170mm \epsffile{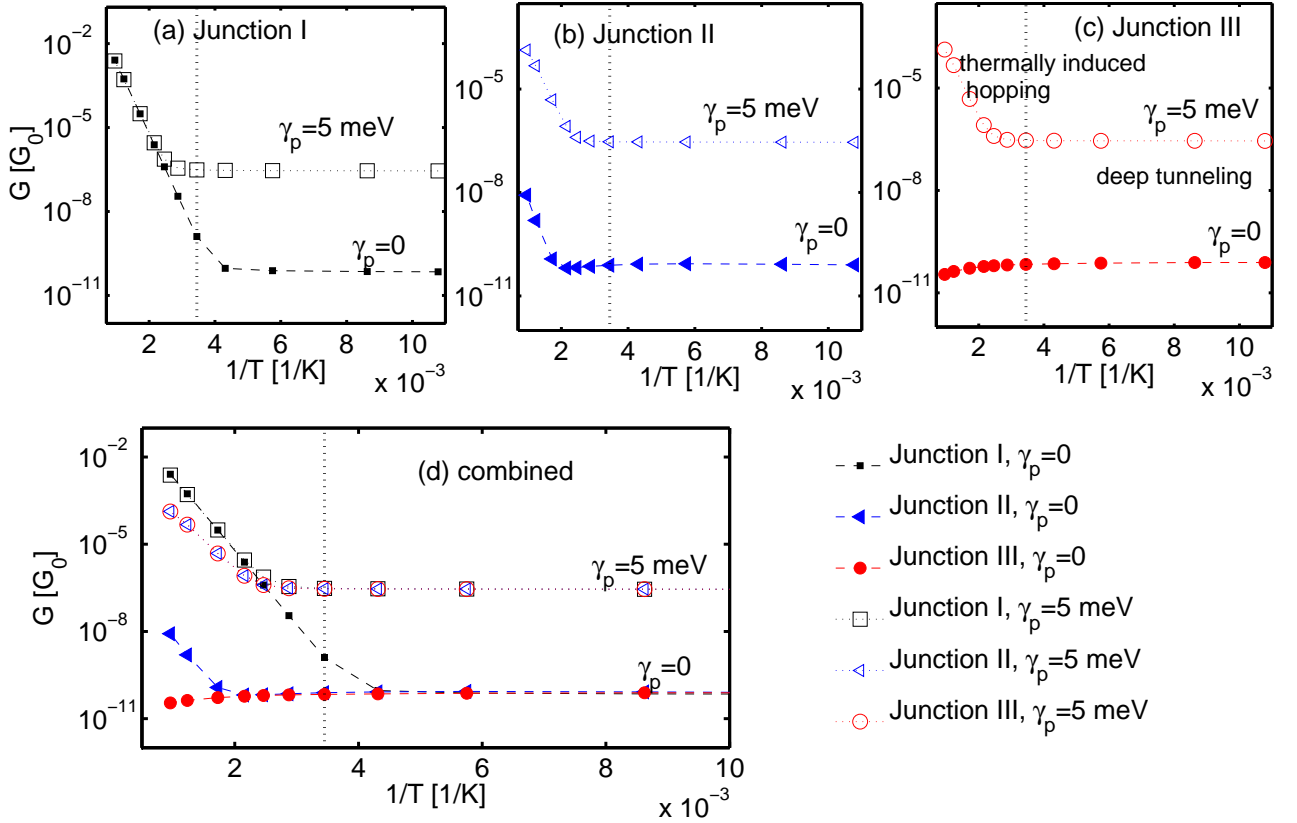}}}
\caption{Temperature dependence of the conductance for
$\gamma_p=0$ meV (full symbols) and
$\gamma_p= 5$ meV (empty symbols).  
(a) Junction I, (b) Junction II, and (c) Junction III. Data from these three panels is combined in panel (d).
We use $\epsilon_B=0.6$ eV, $\epsilon_s=0$, $v=0.05$ eV, $\gamma_{L,R}=0.1$ eV, $n=5$.
The dotted line identifies room temperature.
}
\label{Fig4}
\end{figure*}

Figs. \ref{Fig3}-\ref{Fig4}  display one of the main results of our work: 
 In Junctions II and III,
a thermally activated behavior distinctly identifies incoherent hopping conduction. In contrast, in Junction I, it often 
results from the thermal broadening of the Fermi distribution function, reflecting a coherent transport behavior.

%
In the absence of environmental effects, Fig. \ref{Fig3}, we show that  we can suppress the activated-coherent resonant  behavior using a junction with spacers, for $n=3,5$. 
In longer junctions, $n=7$, the resonant behavior becomes influential even in Junction III
given the negligible contribution of the deep-tunneling current.

In Fig. \ref{Fig4} we analyze the temperature dependence of the conductance  for $\gamma_p\neq0$,
for the three families of molecules. In panel (a), we show that
around room temperature, Junction I  displays a strong thermally activated behavior, with or without environmental effects.
This behavior obviously baffles the identification of the transport mechanism.
%
In junctions III, in contrast,  an activated behavior takes place 
{\it only} for nonzero $\gamma_p$,  see panel (c). 
In panel (b) we show that in Junction II, resonant conduction is well suppressed at room temperature, 
but at very high temperatures, it takes control over the tunneling current. 
We can further suppress the resonant current in Junction II  by increasing the number of spacer units, from two to three.

Other interesting observations in Fig. \ref{Fig4} are that in Junction I,
the thermally activated behavior is delayed to higher temperatures 
when $\gamma_p$ =5 meV, compared to the  $\gamma_p$ =0 case, suggesting that (modified) 
deep-tunneling conductance is enhanced
at finite $\gamma_p$, see analytical results in Sec.  \ref{anal}.
We also find that in the hopping regime, activation energies take similar values,
$E_A=0.50$ eV for Junction I and $E_A=0.37$ eV for Junction II and III.  This
is expected since we construct the spacers so as to minimally affect the energetic of the molecular block.

To summarize our numerical observations, Figs. \ref{Fig1}-\ref{Fig4}: By incorporating spacers at the boundaries of the 
molecule we significantly suppress the resonant tunneling current. 
As a result, we can definitely attribute length and temperature turnover behavior to transition in transport mechanisms,
from tunneling to thermally activated hopping. In the next section we support 
these conclusions with simple analytical considerations.

\subsection{Analytic results}
\label{anal}

We demonstrate here, through a simple calculation, that when spacers are included (Junctions II and III),
the temperature dependence of the conductance predominantly reflects on environmental-inelastic interactions. 
In contrast, without spacers, activated behavior arises predominantly from phase-coherent resonant conduction.

Our minimal model includes a single electronic level of energy $\epsilon_B$, representing the relevant (e.g. LUMO) 
molecular state.
We work in the regime  $\gamma_{L,R}, \beta^{-1}\ll\epsilon_B$.  Energies are set relative to the Fermi energy $\epsilon_F$.
To incorporate the effect of spacers, without including them explicitly, 
we assume here that the metal-molecule hybridization function $\gamma_{L,R}(\epsilon)$ depends on energy:  it is
sharply peaked around the Fermi energy.

To capture the effect of electron-vibration interaction in the molecule, we attach a voltage probe to 
the single molecular level with strength $\gamma_p$.
Following the B\"uttiker's probe technique, 
we enforce the current flowing towards the probe to be zero, 
\bea
&&0=I_p=
\nonumber\\
&& \int \mathcal{T}_{L,P}(\epsilon)[f_L(\epsilon)-f_p(\epsilon)] d\epsilon
+ \int \mathcal{T}_{R,P}(\epsilon)[f_R(\epsilon)-f_p(\epsilon)] d\epsilon.
\nonumber\\
\eea
This condition determines the chemical potential of the probe.
In linear response, setting $\mu_{L,R}=\pm\Delta \mu/2$,  we receive
\bea
&&\frac{\Delta \mu}{2} \int \left[\mathcal{T}_{L,P}(\epsilon) - \mathcal{T}_{R,P}(\epsilon) \right]\left[-\frac{\partial f}{\partial \epsilon}\right] d\epsilon
\nonumber\\
&&= \mu_p \int \left[\mathcal{T}_{L,P}(\epsilon) + \mathcal{T}_{R,P}(\epsilon) \right]\left[-\frac{\partial f}{\partial \epsilon}\right] d\epsilon.
\eea
For simplicity, we assume that the junction is spatially symmetric and introduce the notation
$\gamma_{\nu}(\epsilon)=\gamma_{L,R}(\epsilon)$. Since
$\mathcal{T}_{L,P}(\epsilon)=\mathcal{T}_{R,P}(\epsilon)$, we immediately conclude that the chemical potential
of the probe sits precisely at the center of the bias window,  $\mu_p=0$.

Next, we  calculate the net charge current flowing in the junction, $L$ to $R$,
\bea
I_L&=&\frac{e}{h}\int\mathcal{T}_{L,R}(\epsilon) [f_L(\epsilon)-f_R(\epsilon)] d\epsilon
\nonumber\\
&+& \frac{e}{h}\int \mathcal{T}_{L,P}(\epsilon) [f_L(\epsilon)-f_p(\epsilon)] d\epsilon. 
\label{eq:ILM}
\eea
In linear response, we identify two contributions to the conductance (in units of $G_0$),
\bea
G= G_{coh}+G_{\gamma_p},
\eea
where
\bea
G_{coh}&=&\int \mathcal{T}_{L,R}(\epsilon) \left[-\frac{\partial f}{\partial \epsilon}\right] d\epsilon,
\nonumber\\
G_{\gamma_p}&=&
 \frac{1}{2}\int \mathcal{T}_{L,P}(\epsilon) \left[-\frac{\partial f}{\partial \epsilon}\right] d\epsilon. 
\label{eq:Gsum}
\eea
The transmission functions can be readily calculated \cite{nitzan}, 
\bea
\mathcal T_{L,R}(\epsilon)&=& \frac{\gamma_L(\epsilon)\gamma_R(\epsilon)}{ (\epsilon-\epsilon_B)^2
+ \left[\gamma_L(\epsilon) + \gamma_R(\epsilon)+\gamma_p\right]^2/4},
\nonumber\\
\mathcal T_{L,P}(\epsilon)&=& \frac{\gamma_{\nu}(\epsilon)\gamma_p}{ (\epsilon-\epsilon_B)^2
+ \left[\gamma_L(\epsilon) + \gamma_R(\epsilon)+\gamma_p\right]^2/4}.
\label{eq:TM}
\eea
Recall that the junction is symmetric, $\gamma_{\nu}(\epsilon)\equiv\gamma_{L,R}(\epsilon)$.

Focusing first on the phase-coherent contribution,
$G_{coh}$, we note that the integrand includes two contributions, centered about $\epsilon_B$ and $\epsilon_F$:
First, the derivative of the Fermi function is peaked about the Fermi energy. 
Second, the transmission function is peaked at the energy of the resonant level $\epsilon_B$, as well as around 
$\epsilon_F$---given the special form assumed for $\gamma_{\nu}(\epsilon)$.
As a result, $G_{coh}$ can be approximated by
\bea
G_{coh}= G_{coh}^{deep} + G_{coh}^{res},
\eea
describing off-resonant (deep) and on-resonant (res) contributions, respectively,
\bea
G_{coh}^{deep}&=&
 \frac{\gamma_L(\epsilon_F)\gamma_R(\epsilon_F)}{(\epsilon_F-\epsilon_B)^2+
[\gamma_L(\epsilon_F)+\gamma_R(\epsilon_F)+\gamma_p]^2/4} 
\nonumber\\
G_{coh}^{res}&=& A(\epsilon_B) \frac{\partial f}{\partial \epsilon} \Big|_{\epsilon_B}.
\label{eq:Gcoh}
\eea
Here,
$A(\epsilon)=\pi\frac{\gamma_L(\epsilon)\gamma_R(\epsilon)}{\gamma_L(\epsilon)+\gamma_R(\epsilon)+\gamma_p}$.
Which term dominates? 
If $\gamma_{\nu}$ is assumed a constant---independent of energy---
one could satisfy $G_{coh}^{res}>G_{coh}^{deep}$ at high enough temperatures.
Particularly, in long molecules, the resonant term evolves into a long-range distance-independent conduction,
while $G_{coh}^{deep}$ unfolds into the superexchange behavior, decaying exponentially with distance.
The spacers however shape the density of states, thus the hybridization function, so as
$\gamma_{\nu}(\epsilon_B)\ll \gamma_{\nu}(\epsilon_F)$. As a result,
the resonant current is greatly suppressed, to satisfy $G_{coh}^{res}\ll G_{coh}^{deep}$.

Next, we analyze the incoherent (bath-assisted, $\gamma_p$) contribution.
We again note that since the integrand in Eq. (\ref{eq:Gsum}) is peaked about two energies, 
$\epsilon_{B}$ and $\epsilon_F$, the conductance can be approximated by the sum of two terms,
\bea
G_{\gamma_p}= G_{\gamma_p}^{deep}+G_{\gamma_p}^{T}
\eea
with 
\bea
G_{\gamma_p}^{deep}&=&
\frac{\gamma_{\nu}(\epsilon_F)\gamma_p}{(\epsilon_F-\epsilon_B)^2+[\gamma_L(\epsilon_F)+\gamma_R(\epsilon_F)+\gamma_p]^2/4},
\nonumber\\
G_{\gamma_p}^{T}&=& B(\epsilon_B) \frac{\partial f}{\partial \epsilon} \Big|_{\epsilon_B}.
\label{eq:Ggp}
\eea
Here, $B(\epsilon)=\pi\frac{\gamma_{\nu}(\epsilon)\gamma_p}{\gamma_L(\epsilon)+\gamma_R(\epsilon)+\gamma_p}$.
The first term,
$G_{\gamma_p}^{deep}$, corrects the deep-tunneling conductance of Eq. (\ref{eq:Gcoh}). 
The second term,  $G_{\gamma_p}^{T}$, depends on  the temperature {\it and} on the interaction of electrons to 
the thermal bath, $\gamma_p$.
Since we assume that the temperature is rather low, $\beta\epsilon_B\gg 1$, we receive
an Arrhenius-like form  
\bea
G_{\gamma_p}^{T}
\sim B(\epsilon_B) \beta e^{-\beta\epsilon_B}.
\label{eq:GgpT}
\eea 
In the small coupling limit, $\gamma_p\ll \gamma_{\nu}(\epsilon_B)$,
$G_{\gamma_p}^{T}\propto \gamma_p$. In the opposite limit, the conductance saturates at large $\gamma_p$.

Lets examine now the activated (temperature dependence) behavior showing up in Eqs. (\ref{eq:Gcoh}) and (\ref{eq:Ggp}).
The activated behavior of $G_{coh}^{res}$
arises from the thermal broadening of the Fermi distribution in the leads. 
In contrast, the temperature dependence of $G_{\gamma_p}^T$ unfolds  due to the
coupling of conducting electrons to the probe: 
This channel describes electrons leaving the molecular state to the probe, and 
coming back---potentially with a different phase and energy. 
This scattering event mimics the interaction of electrons with
vibrations, and the thermal factor in Eq. (\ref{eq:Ggp}) thus corresponds to 
environmentally-thermally activated electrons.

We can now organize a simple inequality, 
a condition for thermally induced hopping conducting to take over resonant transport:
%
\bea
B(\epsilon_B) > A(\epsilon_B)  \,\,\,\, {\rm or} \,\,\,\, \gamma_{p}>\gamma_{\nu}(\epsilon_B).
\label{eq:cond}
\eea 
Without spacers, 
the metal-molecule hybridization is typically stronger
than the energy scale for inelastic processes, and it is thus difficult to fulfill this inequality.
In contrast, the effect of spacers translates to a structured density of states at the leads. 
Specifically, in our design, $\gamma_{\nu}(\epsilon_B)\ll \gamma_{\nu}(\epsilon_F)$,
and we are able to satisfy the condition (\ref{eq:cond}).

\begin{figure*}[htbp]
\vspace{0mm} \hspace{-45mm}
{\hbox{\epsfxsize=190mm \epsffile{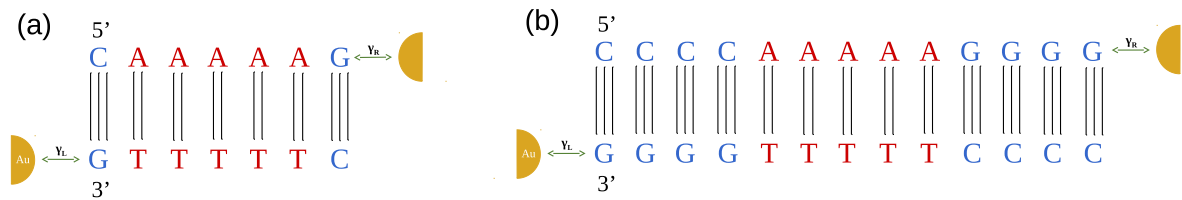}}}
\caption{Examples of DNA Junctions with an A:T block 5 bases long.
(a) Junction II with a single G:C unit at the boundary.
(b) Junction III with a fixed length of $2k+n$=13 bases, (C)$_k$(A)$_{n}$(G)$_k$ with $k=4$ and $n=5$.
}
\label{DNAseq}
\end{figure*}


\begin{table*}[hbtp]
\hspace{10.4mm } Table II:   {\bf  Sequences studied in Fig. \ref{FigDNA1}.}  \\
\begin{tabular}{c c c }
\hline
\hline
Length $n$ of the A:T segment   \,\,\,\, &   5'-C(A)$_n$G-3'  \hspace{8mm }    & 5'-(C)$_k$(A)$_n$(G)$_{k}$-3'
\\ [0.5ex]
\hspace{2.4mm }  (barrier length) &  (Junction II)  \hspace{1mm}  &  (Junction III)\\ [0.5ex]
\hline
 \hspace{11mm } 1 \hspace{14.9mm} & CAG  &  (C)$_6$A(G)$_6$ \\
\hline
\hspace{10mm } 2 \hspace{14.3mm}  &  CAAG  &  \\
\hline
\hspace{10mm } 3 \hspace{14.3mm}  & C(A)$_3$G  &     (C)$_5$(A)$_3$(G)$_5$ \\
\hline
\hspace{10mm } 4 \hspace{14.3mm}  &   C(A)$_4$G    &       \\
\hline
\hspace{10.7mm }5 \hspace{14.7mm}  &   C(A)$_5$G  &         (C)$_4$(A)$_5$(G)$_4$   \\
\hline
\hspace{10.7mm }6 \hspace{14.7mm}  &     C(A)$_6$G   &       \\
\hline
\hspace{10.5mm }7 \hspace{14.7mm}  &    C(A)$_7$G  &     (C)$_3$(A)$_7$(G)$_3$ \\
\hline
\hspace{10.5mm }8 \hspace{14.7mm}  &     C(A)$_8$G   &    \\
\hline
\hspace{10.5mm }9 \hspace{14.7mm}  &    C(A)$_9$G  &     (C)$_2$(A)$_9$(G)$_2$ \\
\hline
\hspace{10.5mm }10 \hspace{14.7mm}  &     C(A)$_{10}$G   &    \\
\hline
\hspace{10.5mm }11 \hspace{14.7mm}  &    C(A)$_{11}$G  &     C(A)$_{11}$G \\
\hline
\hline
\end{tabular}
\label{table:seq}
\end{table*}

\begin{figure}[htbp]
\vspace{0mm} \hspace{3mm}
{\hbox{\epsfxsize=95mm \epsffile{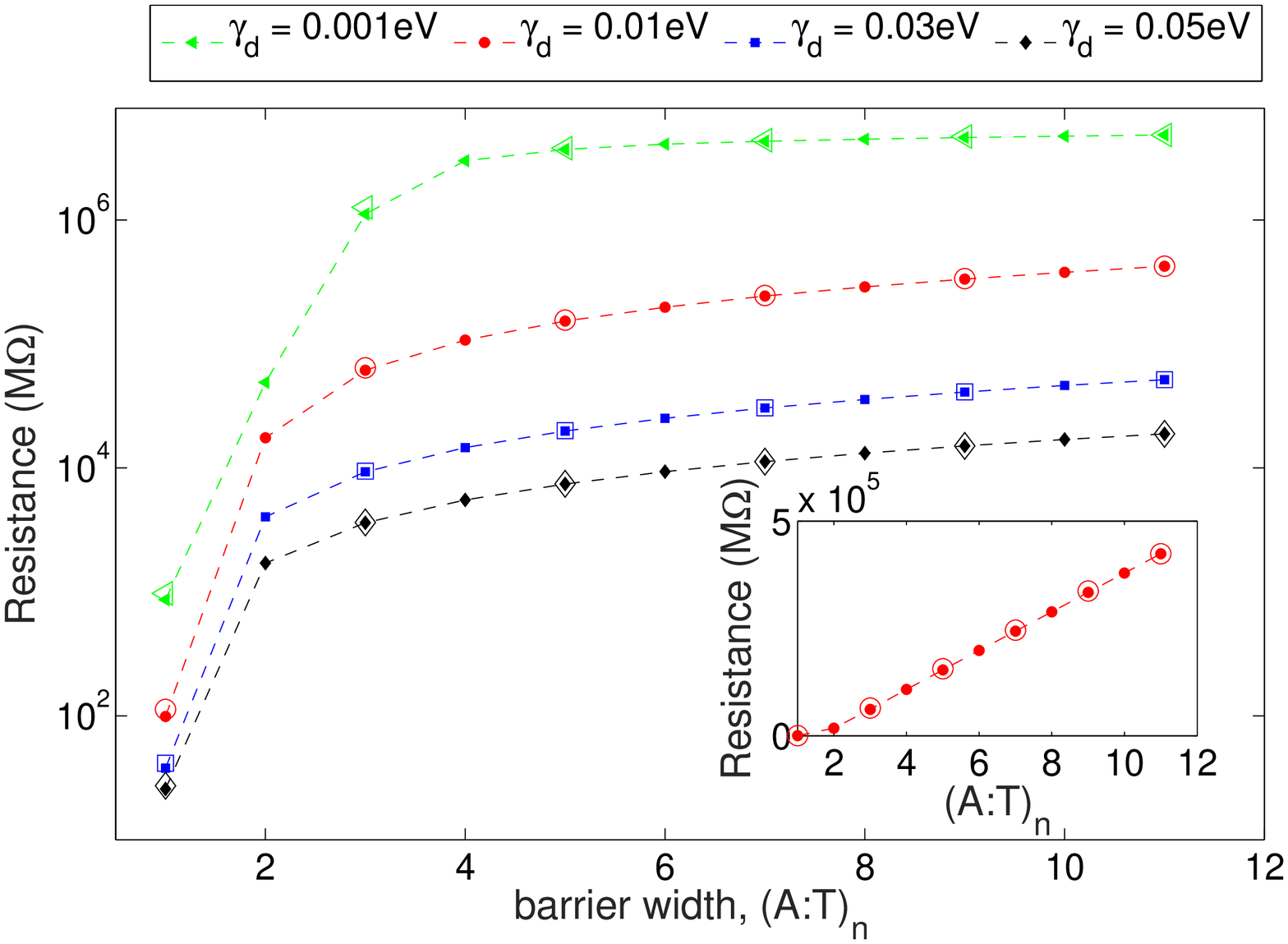}}}
\caption{
Length dependence of the electrical resistance in DNA junctions, Table II.
Junction II (full symbols)  and Junction III  (empty symbols),
under varying degrees of environmental effects, with
$T=295$ K, $\gamma_L = \gamma_R = 0.05$ eV.  $\epsilon_F$ is set at the energy of the G base.
The inset exemplifies the linear dependence of the resistance in the incoherent-hopping regime.
}
\label{FigDNA1}
\end{figure}

\section{DNA molecules with an A:T block}
\label{ResultDNA}

%
Charge transport in DNA sensitively depends on the molecular structure (sequence, length) and its dynamics \cite{Barton10}.
Photochemical and transport measurements 
had indicated that a short A:T block,
embedded within a a G:C sequence,
acts as a tunneling barrier, while the conductance of a longer A:T segment only weakly depends on length  \cite{GieseE,GieseR,janes,Tao04, Tao16}.
This observation indicates on a transition of the charge transport mechanism,
from deep tunneling to thermally-activated hopping conduction.
Tunneling through an A:T barrier has been reported as well in Ref. \cite{janes}, using
a fixed-size junction. In this experiment, a 15 base-pairs DNA molecule was trapped in a small gap, with
the center of the sequence gradually replaced, from G:C to A:T pairs.

In this section, we perform transport calculations for different DNA sequences, 
all with (A:T)$_n$ base-pairs at the center, and (G:C)$_k$ as spacers.
We follow the tunneling-to-hopping crossover as a function of length $n$. 
Our objective is to demonstrate that this turnover behavior
is robust against the number of spacers at the boundaries. This indicates  
that the tunneling-to-hopping crossover observed corresponds 
to the characteristics of the A:T block, rather than the G:C spacers.

We model charge transport in DNA with a tight-binding ladder-model Hamiltonian, 
see e.g. Refs. \cite{cunibertiphonon, ladder1,ladder2,ladder3,BerlinJacs,Wolf,William,Roman}.
This Hamiltonian describes the topology of a double stranded (ds) DNA molecule which is $l$ base-pairs long,
with each site representing a particular base; $L=2l$ is the total number of bases.
We assume that charge transport takes place along the base-pair stacking, and we ignore states on the sugar backbone,
\bea
\hat H_M&=&\sum_{j=1}^n \Bigg[
\sum_{s=1,2} \epsilon_{j,s}\hat c_{j,s}^{\dagger}\hat c_{j,s} 
+ \sum_{s\neq s'=1,2} t_{j,ss'}\hat c_{j,s}^{\dagger}\hat c_{j,s'}
\nonumber\\
&+&
 \sum_{s,s'=1,2} t_{j,j+1,ss'}(\hat c_{j,s}^{\dagger}\hat c_{j+1,s'} + h.c.) \Bigg].
\eea
%
The index $s=1,2$ identifies the strand. $\hat c_{j,s}^{\dagger}$ creates a hole on strand $s$
at the $j$th site with an on-site energy $\epsilon_{j,s}$. $t_{j,ss'}$ and $t_{j,j+1,ss'}$
are the electronic matrix elements between nearest neighboring bases.
This model mimics the topology of ds-DNA: helical effects are
taken into account within renormalized electronic parameters.

We use the parametrization developed in Ref. \cite{BerlinJacs}, 
distinguishing between backbone orientations (5' and 3').
All parameters, including $\epsilon_{j,s}$, $t_{j,ss'}$ and $t_{j,j+1,ss'}$,
are listed in Ref. \cite{BerlinJacs}.
Following Ref. \cite{Wolf}, we simplify this description, and assign a single value (averaged)
for on-site energies for each base, $\epsilon_G=8.178$, 
$\epsilon_A=8.631$ $\epsilon_C= 9.722$ $\epsilon_T=  9.464$, all in eV.
%
We connect the DNA molecule to metal leads as sketched in Fig. \ref{DNAseq},
oriented so as to correspond to experiments \cite{janes}. 

So far, our description concerns the molecular electronic structure of the ds-DNA, assuming 
a rigid structure. To include
environmental effects (structural motion, solvent and counterions dynamical effects) we attach B\"uttiker probes 
to each base. 
Three parameters should be provided as input for the LBP equations:
the position of the Fermi energy $\epsilon_F$ relative to the molecular states,
the metal-molecule hybridization energy $\gamma_{L,R}$, and the electron-environment interaction energy,
encapsulated within $\gamma_p$.
In principle, we could  capture the susceptibility of different bases
and sites along the DNA to environmental interactions by adjusting $\gamma_p$ on each base \cite{Qi}.
Here, for simplicity, we use a single parameter,
 $\gamma_p$, identical for all bases and sites.
For the Fermi energy, we assume that it is set at the energy of the G base,
 $\epsilon_F=\epsilon_G$.
For  $\gamma_{L,R}$, we test values reasonable for molecular junctions, $\gamma_{L,R}=0.05-0.5$ eV. 
In simulations presented below, the conductance was calculated including
both spin species, using $G_0=2e^2/h$.

We study two families of molecules with an A:T block and G:C bases at the boundaries, Junction II and Junction III, 
see examples in Fig. \ref{DNAseq} and a complete list in Table II.
In both cases, we confirm in Fig. \ref{FigDNA1} that the size of the G:C block
does not affect the electrical conductance of the system. Specifically, the tunneling-to-hopping crossover is intact.
This result is significant: It evinces that the conductance measured in the setup
reflects the properties of the A:T bridge, while the spacers play only a small role in the
transport behavior.
In particular, the resistance per site in the hopping regime (inset) is independent of the length of the G:C block.

Several theoretical studies had toyed with the idea of sustaining
coherent resonant transport through long rigid-ordered DNA molecules, achieved through states delocalized over
the bridge \cite{cuniberti-bal,peskin10,peskin12,peskin16}.
In  Fig. \ref{FigDNA1}  we demonstrate that by reducing environmental effects below other energy scales,
$\gamma_p=1$ meV, we reach such a regime, with a very-weak distance dependence for $n>3$, alluding to resonant conduction.
However, other studies of DNA conductance (with the LBP method)
\cite{Qi,William,Roman} point that this parameter should be taken in the range 
$\gamma_p\sim 5-30$ meV, resulting in  hopping conduction for $n\geq 3$.

We summarize our observations:
(i) Our calculations predict tunneling-to-hopping crossover in DNA sequences with a barrier made from an A:T segment.
(ii) Spacers made from G:C base-pairs do not interfere with this crossover behavior.
(iii) The tunneling behavior is quickly washed out beyond $n=3$, turning over to a weak, ohmic distance dependence.
These general observations are in line with experimental work \cite{GieseE,janes,Tao16}. 

\section{Summary} 
\label{Summ}

Distance independent electron transfer, the result of electron transmission through delocalized orbitals, 
is appealing for some molecular electronic applications. However,  resonant conduction
does not reveal information over microscopic electron-nuclei interaction effects within the molecule.
To resolve such phenomena, in this paper we had suggested to suppress the
contribution of resonant electrons by forming molecular junctions 
with extended spacers (anchor groups) at the metal-molecule interface,
designed so as to filter out off-resonance electrons.

Our analytical calculations and numerical simulations had demonstrated the 
adequately and robustness of our approach. 
Specifically, in the presence of spacer groups, thermally activated conduction can be safely attributed 
to the hopping mechanism, rather than to a thermal broadening of the Fermi distribution of the metals.
Similarly, the length dependence of the conductance can be linked to either the tunneling or 
the hopping mechanisms, excluding resonant transmission.
Calculations on DNA junctions with an A:T block had further indicated that the length of the spacer,
a segment of G:C base-pairs,
affects neither the tunneling-to-hopping crossover, nor the resistance per site in the ohmic-hopping regime.

It is interesting to simulate relevant junctions with genuine electron-vibration interaction effects,
using  a first-principle theory \cite{Brandbyge},
and study more deeply the role of extended spacers in transport junctions.
One may also examine the unfurling, length independent transport behavior,
theoretically suggested in Ref. \cite{peskin16} for related DNA junctions.
Finally, DNA nanojunctions show promise for a broad range of applications, including nonlinear charge and energy transport, 
signaling, and sensing \cite{dubi,Hihaths,DNAbook}. Developing theoretical approaches to simulate, explain, and predict
 such functions is indispensable for making progress in this field.

%

\begin{acknowledgments}
The work was supported by the Natural Sciences and Engineering Research Council of Canada and 
the Canada Research Chair Program.
Hyehwang Kim was supported by an Ontario Student Opportunity Trust Funds Research Scholarship.
\end{acknowledgments}
{}

\newpage

\end{document}